\documentclass[conference]{IEEEtran}
\usepackage{graphicx}
\usepackage{cite}
\usepackage{amsmath}
\usepackage{amsfonts}
\usepackage{amssymb}
\usepackage{subfigure}

\newcommand{\bm}{\mathbf}
\newcommand{\be}{\begin{equation}}
\newcommand{\ee}{\end{equation}}
\newcommand{\bea}{\begin{eqnarray}}
\newcommand{\eea}{\end{eqnarray}}
\newcommand{\x}{{\bm x}}

\newcommand{\bD}{{\bf D}}

\newcommand{\bH}{{\bf H}}
\newcommand{\bt}{{\bf t}}

\newcommand{\h}{{\bf h}}

\newcommand{\bv}{{\bf v}}
\newcommand{\bw}{{\bf w}}

\newcommand{\bs}{{\bf s}}

\newcommand{\balpha}{\mbox{\boldmath{$\alpha$}}}

\title{Pilot Decontamination in CMT-based Massive MIMO Networks}

\author{\normalsize Arman Farhang$^*$, Amir Aminjavaheri$^\dagger$, Nicola Marchetti$^*$, Linda E. Doyle$^*$ and Behrouz Farhang-Boroujeny$^\dagger$  
\\$^*$CTVR / The Telecommunications Research Centre, Trinity College Dublin, Ireland, \\
$^\dagger$ECE Department, University of Utah, USA. \\
Email: \{farhanga, marchetn, ledoyle\}@tcd.ie, \{aminjav, farhang\}@ece.utah.edu}

\begin{document}

\maketitle

\begin{abstract}
Pilot contamination problem in massive MIMO networks operating in time-division duplex (TDD) mode can limit their expected capacity to a great extent. This paper addresses this problem in cosine modulated multitone (CMT) based massive MIMO networks; taking advantage of their so-called blind equalization property. We extend and apply the blind equalization technique from single antenna case to multi-cellular massive MIMO systems and show that it can remove the channel estimation errors (due to pilot contamination effect) without any need for cooperation between different cells or transmission of additional training information. Our numerical results advocate the efficacy of the proposed blind technique in improving the channel estimation accuracy and removal of the residual channel estimation errors caused by the users of the other cells.

\end{abstract}

\section{Introduction}
\label{sec:intro}
Recent proposal of massive MIMO as a candidate for the fifth generation of wireless communication networks (5G) has sparked a great deal of interest among researchers. This is due to the fact that they can greatly increase the capacity of multiuser networks. Massive MIMO is a multiuser technique whose ideas are somewhat similar to the spread-spectrum systems. The spreading gains of different users come from the corresponding channel gains between each mobile terminal (MT) antenna and the base station (BS) antennas. Therefore, a significant processing gain can be achieved by utilizing a massive number of antenna elements at the BS. In other words, the processing gain can arbitrarily grow by increasing the number of antenna elements at the BS. As the pioneering work of Marzetta points out, given perfect channel state information (CSI), as the number of BS antennas, in limit, tends to infinity, the processing gain of the system tends to infinity and, accordingly, the effects of both noise and multiuser interference (MUI) completely fade away \cite{Marzetta2010}. Hence, the network capacity (in theory) can be increased unboundedly by increasing the number of antennas at the BS \cite{Marzetta2010}.  

Motivated by Marzetta's observations, \cite{Marzetta2010}, multiple research groups in recent years have studied a variety of implementation issues related to massive MIMO systems, e.g., \cite{Hoydis2013, Hoydis2012, Jose2011, Ngo2011, Gao2011, Payami2012}. As discussed in \cite{Marzetta2010}, the only limiting factor of the network capacity in the single-cell scenario is the coherence time of the channel which needs to be much larger than the number of MTs in order to facilitate accurate channel estimation and allow for data transmission. However, non-cooperative multi-cellular time-division duplex (TDD) networks suffer from a \textit{pilot contamination} problem which was first reported by Jose et al, \cite{Jose2010}. This is a major factor in limiting the capacity of those networks. Due to the channel reciprocity, in TDD systems, the CSI is obtained at the BS during the uplink transmission. Since the channel coherence time is usually not long enough to allow for utilization of orthogonal pilot sequences in different cells, non-orthogonal pilots of neighboring cells will contaminate the pilots of each other \cite{Jose2010}. Thus, the channel estimates at each BS will contain the channel information of the MTs located in the other cells as well as its own users. As a result, when the BS linearly combines the received signal in order to decode the transmitted symbols of its MTs, it also combines the data symbols of the users of other cells which results in inter-cell interference. The corresponding inter-cell interference does not vanish even when the number of BS antenna elements tends to infinity \cite{Marzetta2010}. 

To tackle the pilot contamination problem, a number of solutions have been proposed in the literature \cite{Ngo2012, Yin2013, Muller2014}. The authors of \cite{Ngo2012} use eigenvalue decomposition of the covariance matrix of the received signal samples for channel estimation and show that the channel can be estimated with a higher accuracy compared with linear estimation techniques. In \cite{Yin2013}, the authors propose a cooperative Bayesian channel estimation technique where they find that the pilot contamination effect can be removed completely if the channel covariance matrices satisfy certain conditions. In a more recent work, M{\"u}ller et al, \cite{Muller2014}, propose a blind technique based on a subspace projection method to mitigate the pilot contamination problem which needs singular value decomposition (SVD) of the received signal matrix.

The authors in \cite{FBMCMassive2014} have proposed filter bank multicarrier (FBMC) in massive MIMO application for the first time. They have reported an interesting property of FBMC in massive MIMO systems called \textit{self-equalization} together with various benefits which positions FBMC as a strong candidate in this application. In this paper, we address the pilot contamination problem in massive MIMO networks using cosine modulated multitone (CMT) viz., a particular form of FBMC. This type of FBMC has a blind equalization capability, \cite{Farhang03}. The major contribution of this paper is to extend the blind equalization capability of CMT to massive MIMO application in order to remove the channel estimation errors caused by contaminated pilots. Based on our observations through numerical results, the proposed blind equalization technique is able to remove the pilot contamination effects in multi-cellular massive MIMO networks and converge towards the optimal linear minimum mean square error (MMSE) performance. It is also worth mentioning that our proposed algorithm does not demand any cooperation among different cells. The benefits of FBMC-based massive MIMO are well articulated in \cite{FBMCMassive2014}. The fact that we can use the inherent blind equalization property of the CMT waveform to solve the pilot contamination problem in CMT-based massive MIMO networks is an additional attractive feature to these networks.

The rest of the paper is organized as follows. The system model that is used in this study is introduced in Section~\ref{sec:SysModel}. The blind equalization capability of CMT is discussed in Section~\ref{sec:PDecontamination} and extended to multi-cellular massive MIMO networks to correct the imperfect channel estimates due to the pilot contamination effect present in those networks. The efficacy of our proposed technique is analyzed in Section~\ref{sec:SimResults} through computer simulations. Finally, the conclusions are drawn in Section~\ref{sec:Conclusion}.

Throughout this paper, we use the following notations. Scalars are represented in regular upper and lower case letters. Vectors are represented by boldface lower case letters. Matrices are represented by boldface upper case letters. The matrix or vector superscript $(\cdot)^{\rm H}$ indicates conjugate transpose. $\Re\{\cdot\}$, $\| \cdot \|$ and ${\rm sign}(\cdot)$ demonstrate the real part, Euclidean norm and signum function, respectively. Finally, ${\rm diag}(\x)$, $\mathbb{E}[\cdot]$ and $(\cdot)^{-1}$ identify a diagonal matrix with diagonal elements of the vector $\x$, expectation and inverse of a matrix.  

\section{System Model}\label{sec:SysModel}
We consider a multi-cellular massive MIMO network consisting of $M>1$ cells and $K$ MTs in each cell. Each MT is equipped with a single transmit and receive antenna, communicating with the BS in a TDD manner. Each BS is equipped with $N\gg K$ transmit/receive antennas that are used to communicate with the $K$ MTs in the cell {\em simultaneously}. We also assume, similar to \cite{Marzetta2010}, multicarrier modulation is used for data transmission. However, we replace orthogonal frequency devision multiplexing (OFDM) modulation by CMT modulation.

\begin{figure}
\centering
\includegraphics[scale=0.4]{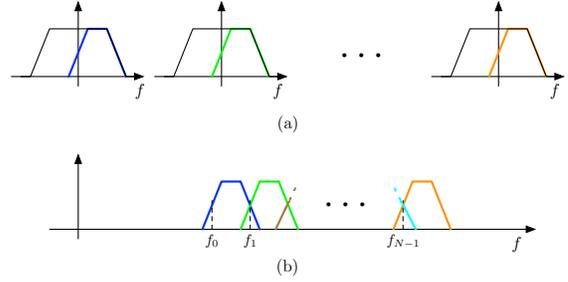}
\vspace{-2 mm}
\caption{CMT modulation. (a) Spectra of baseband data streams (black) and VSB portion of each (other colors). (b) CMT spectrum consisting of modulated versions of the VSB spectra of the baseband data streams to the subcarrier frequencies $f_0$, $f_1$, $\cdots$, $f_{N-1}$.}
\label{fig:CMTmodulation}
\vspace{-5 mm}
\end{figure}

The CMT signal is formed by vestigial side-band (VSB) modulation and upconversion of a set of pulse amplitude modulated (PAM) baseband data streams to different subcarrier frequencies (Fig.~\ref{fig:CMTmodulation}). Moreover, to allow separation of the data symbols (free of inter-symbol interference (ISI) and inter-carrier interference (ICI)), at the receiver, the carrier phase of the VSB signals is toggled between 0 and $\pi/2$ among adjacent subcarriers. The detailed equations explaining why this approach works can be found in \cite{Chang66} and many other publications; a recommended reference is \cite{FarYuen2010}. Reference \cite{FarhangSDRbook} also provides more details, including the implementation structures and their relevant MATLAB codes.

Demodulation of each subcarrier in CMT is a four step procedure.
\begin{enumerate}
\item
For each subcarrier, say, the $k^{\rm th}$ one, the received signal is down-converted to baseband using $f_k$ as the carrier frequency.
\item
The demodulated signal is passed through a matched filter that extracts the desired VSB signal at baseband. The matched filter (MF) removes most of the signal spectra from other subcarriers. However, some residuals of adjacent subcarriers remain which is due to the overlap among the adjacent subcarriers.
\item
The channel effect is removed from the demodulated signal using a complex-valued single tap equalizer. This is based on the assumption that each subcarrier band is sufficiently narrow such that it can be approximated by a flat gain. A multi-tap equalizer may be adopted if this approximation is not valid. 
\item
After equalization, the real part of VSB signal contains the desired PAM symbol only. Its imaginary part consists of a mix of ISI components and ICI components from the two adjacent bands. Accordingly, taking the real part of the equalized VSB signal delivers the desired data signal/symbol, free of ISI and ICI.    
\end{enumerate}

Each MT is distinguished by the BS using the respective subcarrier gains between its antenna and the BS antennas. Ignoring the time and subcarrier indices in our formulation, for simplicity of equations, a transmit symbol $s_{m}(\ell)$ from the $\ell^{\rm th}$ MT located in the $m^{\rm th}$ cell, arrives at the $j^{\rm th}$ BS as a vector 
\be
\x_{j\ell}=t_{m}(\ell)\h_{mj\ell},
\ee
where $t_{m}(\ell)=s_{m}(\ell)+jq_{m}(\ell)$ and $q_{m}(\ell)$ is the contribution of ISI and ICI. $\h_{mj\ell}=[h_{mj\ell}(0),\ldots,h_{mj\ell}(N-1)]^{\rm T}$ indicates the channel gain vector whose elements are the gains between the $\ell^{\rm th}$ MT located in the $m^{\rm th}$ cell and different antennas at the $j^{\rm th}$ BS. The received signal vector at the $j^{\rm th}$ BS, $\x_j$, contains contributions from its own MTs and the ones located in its neighboring cells apart from the channel noise vector $\bv$.
\be\label{eqn:x}
\x_j=\sum_{m=0}^{M-1}\sum_{\ell=0}^{K-1}{\alpha_{mj\ell}\x_{m\ell}} +\bv,
\ee 
where $\alpha_{mj\ell}$'s are the cross-gain factors between the $\ell^{\rm th}$ user of the $m^{\rm th}$ cell and the BS antennas of the $j^{\rm th}$ cell which can be thought as path loss coefficients. In general, $\alpha_{mj\ell}\in[0,1]$. Considering perfect power control for the users of each cell implies that $\alpha_{mj\ell}=1$ for $m=j$. The vector $\x_j$ is fed into a set of linear estimators at the $j^{\rm th}$ BS to estimate the users' data symbols $s_{j}(0)$, $s_{j}(1)$, $\cdots$, $s_{j}(K-1)$. The equation~(\ref{eqn:x}) can be rearranged as
\be\label{eqn:x1}
\x_{j}=\bH_{jj}\bt_j+\sum\limits_{\substack{m=0 \\ m \neq j}}^{M-1}\bH_{mj}\balpha_{mj}\bt_m+\bv,
\ee
where the vector $\bt_m=[t_{m}(0),\ldots,t_{m}(K-1)]^{\rm T}$, $\balpha_{mj}={\rm diag}\{\alpha_{mj0},\ldots,\alpha_{mj(K-1)}\}$ and $\bH_{mj}$'s are $N\times K$ fast fading channel matrices with the columns $\h_{mj\ell},~\ell=0,1,\ldots,K-1$. With the assumption of having the perfect CSI knowledge, the matched filter tap-weight vector for user $\ell$ located in the $j^{\rm th}$ cell can be represented as
\be\label{eqn:MF}
\bw_{j\ell}=\frac{\h_{jj\ell}}{\h_{jj\ell}^{\rm H}\h_{jj\ell}}.
\ee 
The estimated users' data symbols at the output of the matched filters of the cell $j$ can be mathematically written as
\be\label{eqn:MFoutput}
\hat\bs_{j}=\Re\{\bD^{-1}\bH_{jj}^{\rm H}\x_{j}\},
\ee 
where $\bD={\rm diag}\{\|{\h_{jj0}}\|^2,\ldots,\|{\h_{jj(K-1)}}\|^2\}$ and $\hat\bs_j$ is the estimation of the vector $\bs_j=[s_j(0),\ldots,s_j(K-1)]^{\rm T}$ which contains the users' transmitted data symbols. As discussed in \cite{Marzetta2010}, given the perfect CSI knowledge at the BS, when the number of antennas, $N$, tends to infinity, the antenna array gain goes to infinity and hence the multiuser interference and thermal noise effects vanish. As a result, we have $\hat\bs_j=\bs_j$ and the receiver will be optimum. 

However, in practical situations where $N$ is finite, the MF combiner is not optimal and as it is pointed out in \cite{FBMCMassive2014}, MMSE combining has a superior performance which is due to the fact that it maximizes the signal-to-interference-plus-noise ratio (SINR).

The channel gains between the MTs and the BS antennas in each cell are estimated through training pilots transmitted during the uplink phase. The MTs in each cell transmit pilots from a set of mutually orthogonal pilot sequences which allows the BS to distinguish between the channel impulse responces of different users in channel estimation stage. As the authors argued in \cite{Jose2010}, the channel coherence time does not allow the users of neighboring cells to use orthogonal pilot sequences in the multi-cellular scenario. In TDD multi-cellular massive MIMO networks, $M$ base stations use the same set of pilot sequences as well as frequencies. In addition, synchronous transmissions are assumed. Therefore, the same set of pilot sequences being used in neighboring cells will adversely affect the channel estimates at the BS. This effect is called pilot contamination. After correlating the received training symbols with the set of pilot sequences at the BS $j$, the estimates of the channel gains between the MTs and massive array antennas of the BS can be given as
\be\label{eqn:ChanEst}
\hat\bH_{jj} = \bH_{jj}+\sum\limits_{\substack{m=0 \\ m \neq j}}^{M-1}\bH_{mj}\balpha_{mj}+\tilde\bv.
\ee

\begin{figure}
\centering
\includegraphics[scale=0.39]{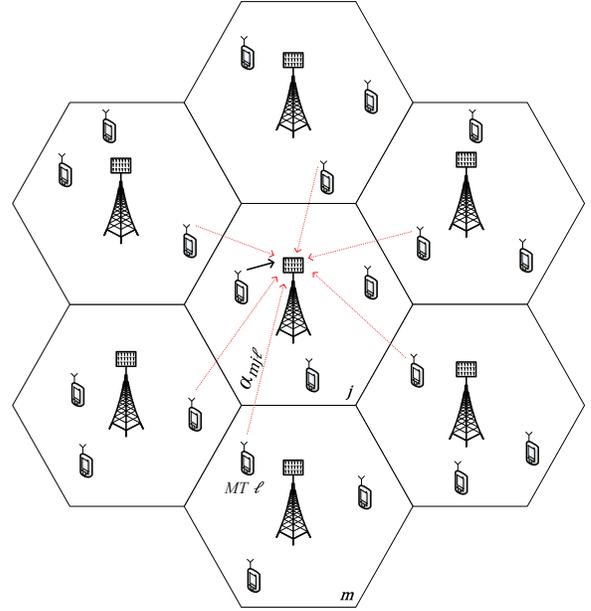}
\vspace{-2 mm}
\caption{Pilot contamination effect in a multi-cellular massive MIMO network.}
\label{fig:Pilot Contamination}
\vspace{-3 mm}
\end{figure}

As one can realize from (\ref{eqn:ChanEst}), the channel estimates at the $j^{\rm th}$ cell are corrupted by the channel impulse responses of its adjacent cells. Therefore, even with infinite number of receive antennas at the BS, there will exist some multiuser interference from the users of other cells. Fig.~\ref{fig:Pilot Contamination} shows this problem in a multi-cellular massive MIMO network where the red arrows show the interference from other cells and the black one shows the transmitted signal of the desired user in the training phase, i.e., uplink transmission. Pilot contamination can have detrimental effects on the performance of multi-cellular networks and greatly impair their sum rate \cite{Marzetta2010}. In the next section, we extend the blind equalization property of CMT to massive MIMO systems in order to purify the channel estimates and tackle the pilot contamination problem without any need for cooperation among the cells or additional training information. 

\section{Proposed Pilot Decontamination Technique}\label{sec:PDecontamination}

As it is noted in \cite{Farhang03}, the imaginary part of the CMT symbol at each subcarrier, i.e. $q_m(\ell)$ is formed from a linear combination of a large number of symbols from the corresponding and also adjacent subcarriers. Following the central limit theorem, one can come up with three observations here:
\begin{enumerate}
\item
The favorable real-part of the equalized CMT symbol at each subcarrier is free of ISI and ICI and so its distribution follows that of the respective PAM alphabet. 
\item
The respective imaginary part suffers from ISI and ICI and is distributed in a Gaussian manner. 
\item 
Both the real and imaginary parts of an unequalized symbol at a subcarrier comprise of ISI and ICI terms and indeed are distributed in a Gaussian manner. 
\end{enumerate}

Based on the aforementioned properties, a blind equalization algorithm similar to the Godard blind equalization algorithm \cite{Godard80} was developed in \cite{Farhang03} such that the cost function

\be\label{eqn:CostFunction}
\xi=\mathbb{E}[(|y_k(n)|^p-R)^2],
\ee
is minimized. $y_k(n)$ is the equalizer output (in the case here, the equalizer output of the $k^{\rm th}$ subcarrier channel), $p$ is integer (usually set equal to 1 or 2), 
$R={\mathbb{E}[|s|^{2p}]}/{\mathbb{E}[|s|^p]}$, 
and $s$ is a random selection from the PAM symbols alphabet.

In this paper, we propose to exploit this algorithm in order to adaptively correct the imperfect channel estimates and hence greatly alleviate the performance degradation due to the contaminated pilots. An LMS (least mean squares) like blind-tracking algorithm, which is computationally inexpensive, based on the cost function (\ref{eqn:CostFunction}) can be adopted. Extension of the proposed blind equalization technique of \cite{Farhang03} to massive MIMO application can be straightforwardly derived as   
\be
\bw_{j\ell}(n+1) = \bw_{j\ell}(n) - 2\mu\hspace{0.5mm}\textrm{sign}(\hat{s}^{(n)}_j(\ell)) (|\hat{s}^{(n)}_j(\ell)|-R)\cdot \x_j(n),
\ee
where $\hat{s}^{(n)}_j(\ell)=\bw_{j\ell}^{\rm H}(n)\x_j(n)$, $\x_j(n)$ is the $n^{\rm th}$ symbol of the received data packet and $\mu$ is an step-size parameter. We initialize the algorithm through the matched filter tap-weight vector
\be\label{eqn:initialization}
\bw_{j\ell}^{(0)}=\frac{\hat{\h}_{jj\ell}}{\hat{\h}_{jj\ell}^{\rm H}\hat{\h}_{jj\ell}},
\ee
where $\hat{\h}_{jj\ell}$ is the estimated channel vector between the user $\ell$ located in the cell $j$ and the $j^{\rm th}$ BS antenna arrays, i.e., the $\ell^{\rm th}$ column of $\hat\bH_{jj}$ in (\ref{eqn:ChanEst}). In the next section, we will show through numerical results that our proposed channel tracking algorithm is able to effectively converge towards the MMSE linear combining with perfect knowledge of channel responses of all the users in all the considered cells, while starting from matched filter tap-weights with imperfect CSI. 
\section{Simulation Results}\label{sec:SimResults}
In this section, we will numerically investigate the performance of our proposed pilot decontamination technique which is presented in Section~\ref{sec:PDecontamination}. This solution extends the blind equalization capability of CMT to massive MIMO networks in order to cope with imperfect channel estimates caused by the pilot contamination effect. 

In our simulations, we consider a massive MIMO network which is comprised of seven cells where the pilot signals of the cell of interest, viz. cell $j$, are interfered by the users of its adjacent cells (Fig.~\ref{fig:Pilot Contamination}). We assume one interferer in each neighbouring cell, i.e., 6 interferers in total whose random cross-gains are less than one. Without loss of generality, in order to expedite our simulations, we consider one user in the $j^{\rm th}$ cell using the same pilot sequence as the users in its neighbouring cells. Thus, its channel estimates at the BS include some residuals from the channel responses of the users of other cells. One transmit antenna is assumed for each user and the number of antennas at the BS are $N=128$. Uncorrelated channel impulse responses between the users and the BS antennas are assumed. The results are for a sample set of channel responses generated based on the Cost 207 channel model for typical urban area with 6 taps. The cross-gain factors, $\alpha_{mj}$'s, are randomly chosen from the range [0,1]. The total bandwidth for this channel is equal to 5 MHz. We take into account $L=256$ number of subcarriers in our simulations whose subcarrier spacing is equal to 19.531 kHz. Binary PAM signaling is used in our simulations. This is equivalent to QPSK (quadrature phase shift keying) signaling if OFDM was adopted. The target SINR in our simulations is set equal to 32 decibels (dB).

After the BS estimates the channel responses of its users through their pilot sequences, the equalizer tap-weights are initialized using equation (\ref{eqn:initialization}). Then filter tap-weight adaptation will be performed. Due to the different channel gains, each subcarrier has a different signal level. Therefore, the step-size $\mu$ will be normalized with respect to the instantaneous signal energy in each iteration. Accordingly, for the binary PAM signaling, the combiner tap-weights can be updated using
\begin{eqnarray}\label{eqn:TapUpdate}
\bw_{j\ell}(n+1) = \bw_{j\ell}(n) - \frac{2\mu}{\x^{\rm H}_j(n)\x_j(n)+\epsilon}\hspace{0.5mm}\textrm{sign}(\hat{s}^{(n)}_j(\ell)) \nonumber \\ \times (|\hat{s}^{(n)}_j(\ell)|-R)\cdot \x_j(n),
\end{eqnarray}
where $\epsilon$ is a small positive constant which assures numerical stability of the algorithm when the term $\x^{\rm H}_j(n)\x_j(n)$ has a small value.

Our proposed pilot decontamination technique is evaluated through looking into its SINR performance and comparing it with the SINRs of the MF and MMSE detectors having the perfect CSI knowledge of all the users in all the cells. The SINRs of the MF and MMSE linear combiners are calculated based on their derived formulas in \cite{FBMCMassive2014}. It is worth mentioning that for the MMSE combining, the $j^{\rm th}$ BS needs to know the perfect channel impulse responses of all the users located in its neighbouring cells having the same pilot sequences as its own user. 

\begin{figure}
\centering
\includegraphics[scale=0.68]{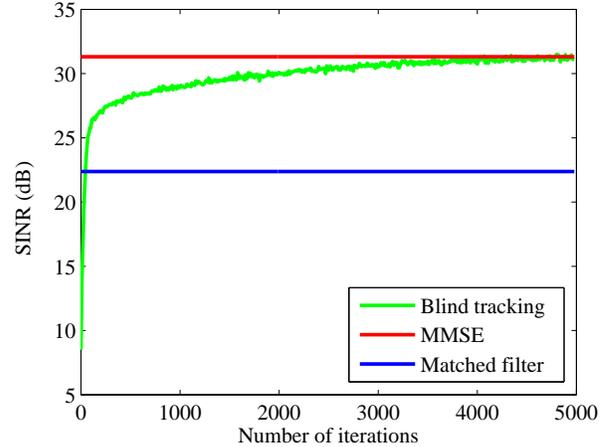}
\vspace{-4 mm}
\caption{SINR comparison of our proposed blind tracking technique with respect to the MF and MMSE detectors having the perfect CSI knowlege.}
\label{fig:SINR}
\vspace{-3 mm}
\end{figure} 
  
Fig.~\ref{fig:SINR}, shows the SINR performance of the proposed blind tracking technique in dB with respect to the number of iterations. As the figure depicts, there exists an abrupt SINR improvement during the first 50 iterations where the output SINR of the blind combiner reaches that of the MF combiner with the perfect CSI knowledge. Running larger numbers of iterations has shown that the output SINR of our blind channel tracking technique can suppress the pilot contamination effect and converge towards that of the MMSE combiner. Apart from its high computational complexity, the MMSE detector needs perfect knowledge of the channel impulse responses between the interfering users of the other cells and its array antennas. This clearly is an impossible condition. Methods proposed here, on the other hand can approach MMSE performance, simply by running an LMS like algorithm. 

\begin{figure}
\centering
\includegraphics[scale=0.6]{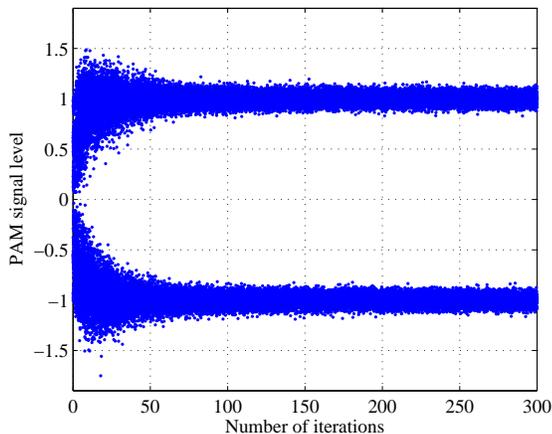}
\vspace{-2 mm}
\caption{Eye pattern of the combined symbols using the proposed blind tracking technique.}
\label{fig:Const}
\vspace{-5 mm}
\end{figure}

Fig.~\ref{fig:Const} represents the eye pattern of the detected symbols using our proposed blind combiner with respect to the number of iterations. As can be seen, the eye pattern of the detected symbols improves as the number iterations increases.

The existing techniques in the literature which address the pilot contamination problem, \cite{Ngo2012, Yin2013, Muller2014}, either need cooperation between the cells or they are computationally intensive. The solutions that are applicable to non-cooperative cellular networks are of interest to this paper \cite{Ngo2012, Muller2014}. The solution proposed in \cite{Ngo2012} needs  eigenvalue decomposition of the covariance of the received signal and the one presented in \cite{Muller2014} has to calculate the SVD of the received signal matrix. The involved matrices are of size $N$ (128 for the examples given here). However, our solution simply needs to update the combiner tap-weights using (\ref{eqn:TapUpdate}). Apart from computational complexity, the pilot decontamination method that is proposed in this paper is also structurally much simpler, as no matrix operation is involved.

From Fig.~\ref{fig:SINR}, one may realize that a large number of iterations are needed for our algorithm to approach the SINR performance of the MMSE detector. This does not mean that a very long packet of data is needed for this algorithm to converge towards the MMSE detector's performance. The MMSE performance can be achieved through multiple runs of the algorithm over a much shorter packet of data. 

\section{Conclusion}\label{sec:Conclusion}
In this paper, we addressed the pilot contamination problem in a TDD multi-cellular massive MIMO network. The pilot contamination problem can adversely affect the performance of the massive MIMO networks and as a result create a great amount of multi-cell interference in both uplink and downlink transmissions. The blind equalization capability of CMT is extended to massive MIMO networks to mitigate the pilot contamination effect. The performance of our proposed solution is analysed through computer simulations. It has been shown that starting from corrupted channel estimates, after running a small number of iterations our algorithm performs as well as or better than the matched filter with perfect CSI. We have shown that the output SINR of our algorithm converges towards that of the MMSE solution where the BS should have the perfect knowledge of CSI for all the users located in its neighbouring cells. 

\bibliographystyle{IEEEtran}

\end{document}